\documentclass[a4paper]{article}

\usepackage{INTERSPEECH2021}
\usepackage{multirow, subcaption}
\usepackage{xcolor}
\usepackage{ulem}

\title{An Attribute-Aligned Strategy for Learning Speech Representation}
\name{Yu-Lin Huang$^{1,3}$, Bo-Hao Su$^{1,3}$, Y.-W. Peter Hong$^{1,2,3}$, Chi-Chun Lee$^{1,2,3}$}
\address{
$^{1}$Department of Electrical Engineering, National Tsing Hua University, Taiwan \\
$^{2}$Institute of Communication Engineering, National Tsing Hua University, Taiwan\\
$^{3}$MOST Joint Research Center for AI Technology and All Vista Healthcare, Taiwan}
\email{
huang8592301@gapp.nthu.edu.tw, borrissu@gapp.nthu.edu.tw, \\ 
ywhong@ee.nthu.edu.tw, cclee@ee.nthu.edu.tw}

\begin{document}

\maketitle
\begin{abstract}
  Advancement in speech technology has brought convenience to our life. However, the concern is on the rise as speech signal contains multiple personal attributes, which would lead to either sensitive information leakage or bias toward decision. In this work, we propose an attribute-aligned learning strategy to derive speech representation that can flexibly address these issues by attribute-selection mechanism. Specifically, we propose a layered-representation variational autoencoder (LR-VAE), which factorizes speech representation into attribute-sensitive nodes, to derive an identity-free representation for speech emotion recognition (SER), and an emotionless representation for speaker verification (SV). Our proposed method achieves competitive performances on identity-free SER and a better performance on emotionless SV, comparing to the current state-of-the-art method of using adversarial learning applied on a large emotion corpora, the MSP-Podcast. Also, our proposed learning strategy reduces the model and training process needed to achieve multiple privacy-preserving tasks.
  
\end{abstract}

\section{Introduction}
\label{sec:intro}

Speech technology has rapidly proliferated and integrated deeply into our daily life \cite{akccay2020speech,10.1145/1873951.1874246,xie2019utterance,braga2019automatic,toth2018speech}. While these applications bring convenience to our life, several growing concerns have gained attention and need to be addressed with care. The first is \textit{privacy} due to concerns of sensitive information leakage: for example, users may not expect to disclose their identity information while using a speech emotion recognition (SER) system; on the other hand, users may not wish to share their emotional condition when being assessed by a speaker recognition (SR) system. Moreover, the collective social norm would create unwanted and often detrimental self-exaggerated issues around equality, e.g., unfair biases toward gender types \cite{tatman2017gender} or race \cite{sap2019risk}, when using data-driven approaches for speech technology. Speech is an informative signal which contains personal sensitive attributes by nature; hence developing appropriate methods either to protect privacy information, such as identity and emotion, or to mitigate the undesired biases, like gender and race, is critical in the current era.

Recently, several works in speech processing have started to address these issues using privacy-aware representation learning. For example, Srivastava et al.\ used adversarial representation learning on automatic speech recognition (ASR) to protect speaker identity \cite{Srivastava2019}, Alouf et al.\ used CycleGAN-based method to generate emotion-less synthesized speech for voice assistant to hide personal affect \cite{aloufi2019emotionless}, Jaiswal et al.\ used adversarial learning to generate gender-invariant representation for identify-free emotion recognition \cite{Jaiswal2020PrivacyEM}, and Xia et al.\ applied adversarial learning to mitigate racial bias in hate speech detection \cite{xia-etal-2020-demoting}. While current state-of-the-art methods concentrate on using adversarial learning, this strategy suffers from several shortcomings. Adversarial method address privacy issues by learning a speech signal space with no targeted sensitive attributes as measured by its ability in fooling a well-trained discriminator that is in charge of classifying sensitive information, e.g., gender and speaker identity. This attribute invariant learning strategy lacks a flexible mechanism to adapt to different criterion of privacy preserving; for example, in some tasks only the ``gender'' attribute may need to be protected while some other tasks would require the ``speaker identity" to be private. For different scenario of interest, one would have to re-train the adversarial network over again.

In this work, instead of taking a `per-attribute' adversarial invariant learning approach, we formulate the problem as devising a learning strategy that would result in attribute-aligned speech representation. The core idea centers on conceptualizing that speech contains a mixture of attributes, \cite{Hsu2017,8462169}, e.g., gender, age, emotion and semantics, etc. By factorizing the entangled information of speech signal into independent attributes with proper attribute-alignment, we can protect particular sensitive information by attribute selection, i.e., masking targeted sensitive attributes, to minimize either privacy-leakage or biased decision. In this paper, we evaluate this idea by targeting two sensitive attributes in speech, i.e., emotion and identity, and our aim is to show that this approach can flexibly achieve privacy-preserving applications by eliminating identity contents in SER or emotion contents in SV at ease. 



We propose a framework of flexible attribute masking for speech, inspired by the fair representation learning \cite{pmlr-v97-creager19a}. We aim to learn a layered disentangled speech representation with a backbone of variational autoencoder (VAE) \cite{Kingma2014,DBLP:journals/ftml/KingmaW19}. We specifically propose a layered dropout strategy in a multi-task framework to achieve attribute-alignment, i.e., forces the latent to align in an emotion-related to identity-related order. To further \textit{clean up} the aligned representation knowing that these two attributes are highly correlated \cite{8925044,pappagari2020x}, we add adversarial reversal layer to each task-specific branch. Our strategy provides flexibility in either identity masking or emotion masking to come up with an identity-free latent for privacy-preserving SER or emotionless latent for privacy-preserving SV with a unified learning framework. In this work, we evaluate our method on MSP-Podcast \cite{8003425} for SER and SV tasks using three types of feature, and achieve competitive results on SER (emobase: 52.41\% weighted f-score, 41.14\% EER), and an improvement on SV (netvlad: 34.35\% weighted f-score, 10.91\% EER; x-vector: 34.23\% weighted f-score, 9.63\% EER), compared to the state-of-the-art adversarial learning method.

\begin{figure*}[t]
\centering
\vspace{-9mm}
\includegraphics[width=0.9\linewidth,]{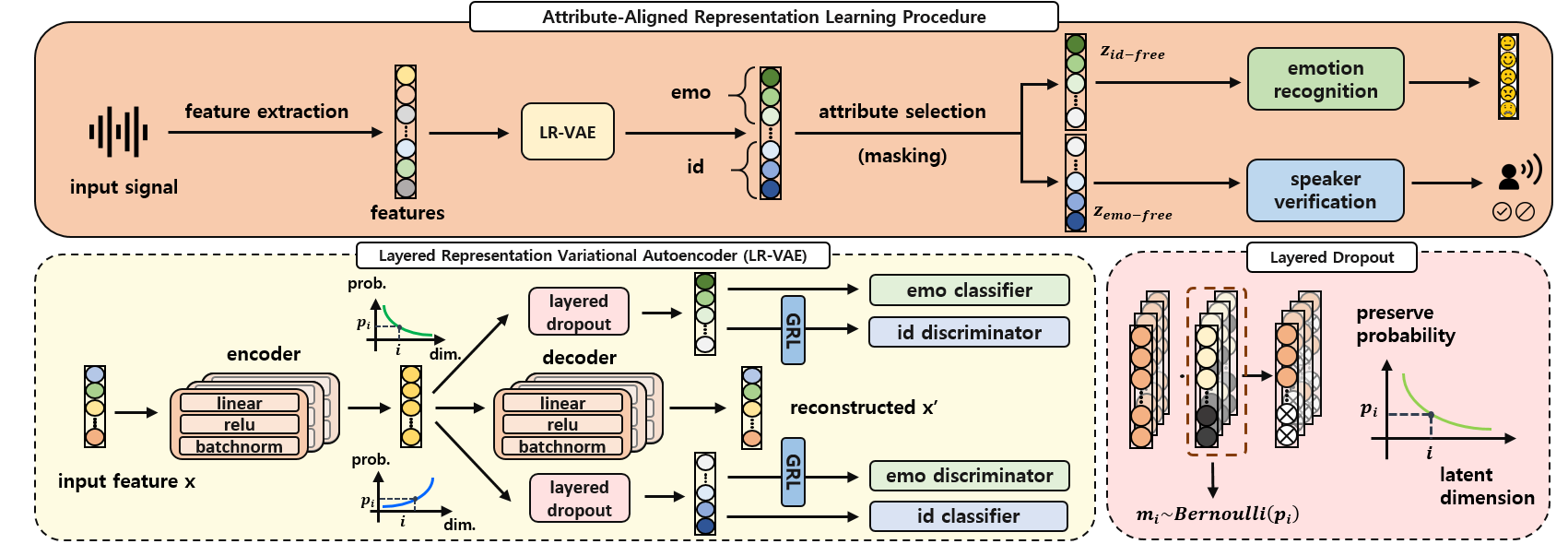} 
\caption{An illustration of our proposed method for attribute-aligned representation learning. It includes three blocks: representation learning procedure, LR-VAE, and layered dropout. Notice that $Z_{id-free}$ stands for identity-free representation, $Z_{emo-free}$ stands for emotionless representation.}
\label{fig:model}
\vspace{-5mm}
\end{figure*}

\section{Methodology}

\subsection{Dataset Description}

In this study, we focus on two main tasks, emotion recognition and speaker verification. To evaluate the performance of these two tasks, a large corpus with emotional labels and multiple speakers is needed. Hence, we use the MSP-Podcast database \cite{8003425}, which includes over 1,000 podcast recordings. Each podcast is segmented into speaking turns, where segments with music, overlapped speech, telephone quality speech and background noise are discarded. 

In this work, we use data with 5 categorical emotions: neutral, angry, sad, happy and disgust as in \cite{pappagari2020x}. We used the standard splits in Release 1.4 for training, development, and testing, which includes 610 speakers in train set, 30 speakers in development set, and 50 speakers in test set, where each set of speakers are disjoint. The distribution of the 5 emotion classes are: angry: 8.81\%, happiness: 27.10\%, neutral: 53.05\%, sad: 3.95\%, disgust: 7.09\%.

\subsection{Feature Extraction}

In this work, we use three different input features for the two tasks: emobase2010, netvlad embedding, and x-vector embedding to verify the effectiveness of our proposed method. First, we use emobase2010, which is a commonly used feature for SER, as input. It is a 1582 dimensional feature including pitch, loudness, mfcc and spectral, etc. We extract emobase2010 using openSMILE toolkit \cite{10.1145/1873951.1874246}. Further, we extract embeddings commonly used in state-of-the-art speaker verification task, i.e., netvlad \cite{xie2019utterance} and x-vector \cite{pappagari2020x}. The netvlad embedding is extracted using the released pre-trained model \cite{xie2019utterance}, while the x-vector embedding is obtained by training on the Voxceleb2 \cite{Chung2018} using the structure mentioned in \cite{pappagari2020x}.

\begin{table*}[th]
  \vspace{-4mm}
  \caption{The experiment results are presented in weighted f-scores (\%) and EER (\%) for SER and SV respectively. Notice that PP stands for privacy-preserving, where columns of origin stand for original representation without privacy protection, PP-SER stands for identity-free SER, and PP-SV stands for emotionless SV.}
  \label{tab:compare}
  \centering
  \begin{tabular}{l|c|c|c|c|c|c|c|c|c|c|c}
    \toprule
    \multicolumn{2}{c|}{\multirow{2}{*}{\textbf{Method}}} &
    \multicolumn{1}{c|}{\textbf{DNN}} &
    \multicolumn{1}{c|}{\textbf{VAE}} &
    \multicolumn{2}{c|}{\textbf{A-VAE}} &
    \multicolumn{3}{c|}{\textbf{LR-VAE (w/o adv)}} &
    \multicolumn{3}{c}{\textbf{LR-VAE}} \\ \cline{3-12}
    \multicolumn{2}{c|}{} & origin & origin & PP-SER & PP-SV & origin & PP-SER & PP-SV & origin & PP-SER & PP-SV \\
    \hline\midrule
    \multirow{2}{*}{\textbf{emobase}} &
    emo & 54.90 & 52.61 & 52.09 & 37.93 & 53.54 & \textbf{52.71} & 40.33 & 52.86 & 52.41 & \textbf{37.54} \\ \cline{2-2}
    & id & 14.45 & 12.99 & \textbf{41.49} & 17.16 & 13.27 & 30.64 & 19.16 & 11.77 & 41.14 & \textbf{12.70} \\
    \midrule
    \multirow{2}{*}{\textbf{netvlad}} & 
    emo & 52.99 & 49.80 & \textbf{49.24} & 38.92 & 50.11 & 49.04 & \textbf{34.10} & 50.01 & 48.23 & 34.35 \\ \cline{2-2}
    & id & 8.14 & 8.37 & 40.66 & 15.25 & 8.20 & 32.39 & 14.51 & 8.53 & \textbf{49.51} & \textbf{10.91} \\
    \midrule
    \multirow{2}{*}{\textbf{x-vector}} &
    emo & 53.24 & 52.60 & 48.95 & \textbf{34.13} & 52.22 & \textbf{51.76} & 36.05 & 52.57 & 51.09 & 34.23 \\ \cline{2-2}
    & id & 10.05 & 8.55 & 45.93 & 13.56 & 8.75 & 37.61 & 10.66 & 8.44 & \textbf{49.48} & \textbf{9.63} \\
    \bottomrule
  \end{tabular}
  \vspace{-4mm}
\end{table*}

\subsection{Layered Representation Variational Autoencoder}

We propose a layered-representation variational autoencoder (LR-VAE) to factorize the entangled dimensions contained in speech and arrange these dimensions in an emotion-related to identity-related order. LR-VAE contains two main components, i.e., disentangled representation and layered dropout. We will first describe VAE, i.e., a well-known structure for disentangled learning. Then, we will further detail our layered dropout with adversarial multitask learning to obtain attribute-aligned speech representation.

\subsubsection{Variational Autoencoder (VAE)}
\label{sssec:vae}

In this work, we use disentangled representation learning via VAE to derive a latent node-wise independent representation. VAE model aims to learn the marginal likelihood of a data $\mathbf{x}$, with the objective function:
\vspace{-0.5mm}
\begin{equation}
\label{vae_loss}
\mathcal{L}_{VAE} = \mathbb{E}_{q(z|x)}[\log p(x|z)]-D_{KL}(q(z|x) || p(z))
\vspace{-0.5mm}
\end{equation}

where $D_{KL}(||)$ stands for the non-negative Kullback-Leibler divergence. The KL-divergence term encourages the posterior distribution to be close to an isotropic Gaussian to achieve disentanglement purpose.





\vspace{-0.5mm}
\subsubsection{Layered Dropout with Adversarial Multitask Learning}
\label{sssec:dropout}

In this work, we propose a strategy of layered dropout with a multitask-learning architecture. Multitask learning aims to include both emotion and speaker identity information into the latent codes. Layered dropout is further utilized to force these attributes to align toward both ends of the latent codes resulting in a layered representation. Also, adversarial branches, i.e., gradient reversal layer, are used to additionally `purify' this attribute-aligned representation.

Dropout is a well-known regularization method in deep learning to prevent neural networks from overfitting\cite{10.5555/2627435.2670313}. Layered dropout works in a similar manner but with a different purpose. We propose to use this as a learning mechanism to make each dimension of the latent codes carry different importance to the designated task. In our work, the two tasks are defined as the emotion recognition and the speaker verification. We design a dropout rate function making the probability of dropping decreases (or increases) monotonically for each node of the input layer. This effectively forces the target task's discriminatory information to concentrate on nodes with lower dropout rates. 

Let $\mathbf{x}$ denotes the input vector with $N$ dimensions of a layer of a neural network, we define a vector with decreasing preserving rates $\mathbf{p}$ for task of emotion recognition (increasing preserving rates for speaker verification), where $0\le p_{i}\le 1$ for $i \in \{0,\dots,N-1\}$. With layered dropout, the input vector $\mathbf{x}$ of the feed-forward operation is replaced by vector $\mathbf{\widetilde{x}}$, generated by:
\vspace{-1mm}
\begin{equation}
m_i \sim \text{Bernoulli}(p_i)
\end{equation}
\begin{equation}
\mathbf{\widetilde{x}} = \mathbf{m} * \mathbf{x}
\end{equation}
Here, $*$ denotes an element-wise product. $\mathbf{m}$ acts as a mask before the vector $\mathbf{x}$ is fed into the layer. For a dimension $m_{i}$ in the vector $\mathbf{m}$, it's an independent Bernoulli random variable with probability $p_{i}$ being 1, which means to preserve the $i^{th}$ node, and 0 means to drop the $i^{th}$ node. While testing, $\mathbf{W}$ represents for weights of network and the weights are scaled as $\mathbf{W}_{test}=\mathbf{W}*\mathbf{p}^\top$ and inference without dropout, which is same as the vanilla dropout layer.

This layered dropout mechanism alters the dropout rates being applied on both sides of the representation before an emotion (identity) classifier, the latent codes form an aligned emotion-to-identity order from top end to bottom end during the optimization step. Furthermore, we add an auxiliary mechanism of adversarial branches with gradient reversal layers \cite{ganin2015unsupervised} during multitask learning. The goal is to learn cleaner factorized identity-free (emotion-free) representation. After having an attribute-aligned representation, we simply need to mask the dimension representing the particular attribute of interest. For example, to protect identity information in SER, we can simply mask the nodes that have high emotion preservation rates and low speaker identity preservation rates, and vice versa for in SV. Notice that our attribute aligned strategy provides a mechanism to select ``what to protect'' with a single unified learning, which is more efficient than the adversarial method that requires re-training in different scenarios.
\vspace{-2mm}
\section{Experiments}

\subsection{Experiment Setup}
\label{ssec:setup}
The structure of our VAE model is as follows: multi-layer perceptron (MLP) is applied for encoder and decoder. Additionally, fully connected layer is applied to model the mean and log variance of the latent code for the encoder. For multi-task learning, two MLP classifiers are trained for emotion recognition and speaker identification, and two MLP discriminators are trained for adversarial learning by applying gradient reversal layers \cite{ganin2015unsupervised}. We set the learning rate as $5e^{-4}$, and the batch size as 128. Moreover, we add a regularization of $1e^{-6}$ to all weights and biases to stabilize the training process. Rectified Linear Unit (ReLU) is chosen as the activation function. We train the model using Adam optimizer, with $L_{obj}$ as the objective, defined as:
\vspace{-1.5mm}
\begin{equation}
L_{obj}=L_{VAE}+L_{emo}+L_{id}+L_{emo-adv}+L_{id-adv}
\vspace{-1.5mm}
\end{equation}
where $L_{VAE}$ represents the reconstruction error and KL divergence loss as defined in equation \ref{vae_loss}, while $L_{emo}$ and $L_{id}$ represents the cross-entropy loss for emotion recognition and speaker identification; $L_{emo-adv}$ and $L_{id-adv}$ represents the adversarial loss for emotion recognition and speaker identification. Notice that for speaker verification (SV) task, models are trained to predict speaker identity in the training set, i.e., speaker identification, to learn identity-related information during training; while during evaluation, the hidden layer embedding is extracted and apply to speaker verification system.

We evaluate the performance of SER using weighted f-score (WFS), following the experiment setup in \cite{pappagari2020x}, and evaluate the performance of speaker verification by equal error rate (EER). For each feature set, we train a factorized layered representation encoder based on training set, select model using validation set, and test performance on test set. Assuming attackers have access to the training set with encoded representation and labels of speakers. Our goal is to generate a representation such that for the encoded representations with particular sensitive attributes masked, neither attackers nor hosts are able to identify the sensitive attributes while the main task performance is maintained.

\begin{figure*}[htb!]
\vspace{-4mm}
\minipage{0.33\textwidth}
\begin{subfigure}{0.5\textwidth}
\centering
\includegraphics[width=1.0\linewidth, width=5.5cm]{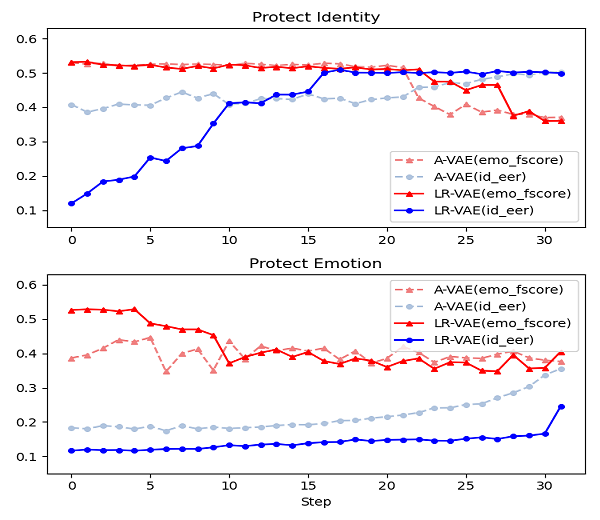}
\caption{emobase}
\label{fig:emobase}
\end{subfigure}
\endminipage
\minipage{0.33\textwidth}
\begin{subfigure}{0.5\textwidth}
\centering
\includegraphics[width=1.0linewidth, width=5.5cm]{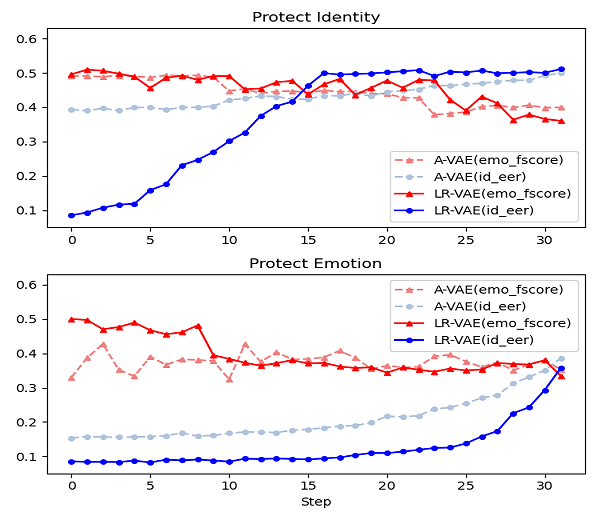}
\caption{netvlad}
\label{fig:netvlad}
\end{subfigure}
\endminipage
\minipage{0.33\textwidth}
\begin{subfigure}{0.5\textwidth}
\centering
\includegraphics[width=1.0\linewidth, width=5.5cm]{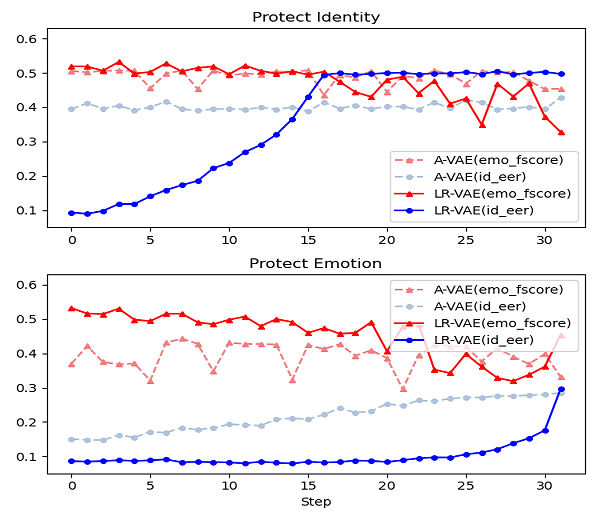} 
\caption{x-vector}
\label{fig:xvector}
\end{subfigure}
\endminipage
\caption{The performance curves in the masking experiment where the y-axis for SER results are weighted f-scores, and EER for SV. For the first row (protect identity), the masking process starts in a bottom up order, while for the second row (protect emotion), we start masking from the top group.}
\label{fig:curve}
\vspace{-5 mm}
\end{figure*}

\subsubsection{Baseline Methods}
\label{sssec:baseline}
The following are the baseline methods of different learning strategies that we use to compare with LR-VAE. Notice that privacy-preserving (PP) on LR-VAE are done by masking the dimension of particular sensitive attributes in the latent codes. \\
{\bf DNN}: A model conducted by fully connected layers to obtain the baseline performance on SER and SV for each feature.\\
{\bf VAE}: A vanilla VAE trained by multi-task learning on SER and SV tasks. \\
{\bf A-VAE}: A VAE trained for single task (SER or SV) with adversarial learning (reverse gradient) on the other task (SV or SER). \\
{\bf LR-VAE (w/o adv)}: A model similar to our proposed LR-VAE, but trained without adding the adversarial branch.

\subsection{Result and Analysis}
\label{ssec:analysis}

\subsubsection{Sensitive Attribute Protection}
\label{sssec:performance}

Note that all the comparison are presented in absolute points in this section. For privacy-preserving speech emotion recognition (PP-SER), we aim to protect user's identity information while preserving the emotion recognition performance. As shown in the PP-SER columns in table \ref{tab:compare}, our proposed LR-VAE achieves the better privacy preserving performance on x-vector and netvlad and similar result on emobase comparing to A-VAE. It shows that our proposed method is able to obtain a competitive emotion recognition performance (0.32\% WFS higher on emobase, 1.01\% WFS lower on netvlad, and 2.14\% WFS higher on x-vector), with better improvements in protecting speaker identity (only 0.35\% EER worse on emobase, 8.85\%  EER better on netvlad, and 3.55\% EER better on x-vector).

On the other hand, to achieve emotion-protected speaker verification (PP-SV), we aim to reduce users' emotional information in the speech while preserving the speaker verification performance. As shown in PP-SV columns in table \ref{tab:compare}, our proposed LR-VAE achieves the best emotion protection performance on all three features comparing to A-VAE. It shows that our proposed method could better maintain the speaker verification performance (4.46\% EER better on emobase, 4.34\% EER better on netvlad, and 3.93\% EER better on x-vector), while achieving state-of-the-art emotion-related attributes protection (0.39\% WFS better on emobase, 4.57\% WFS better on netvlad, and 0.10\% WFS worse on x-vector)

We first study the baseline DNN results shown in the column, DNN, in table \ref{tab:compare}. The promising performance show that regardless of features, it contains both emotion and identity information. It reinforces the current concerns that speech contains many personal attributes that users may not want to reveal. Then, we compare the DNN results to VAE results shown in the column, VAE origin, in table \ref{tab:compare}. We do see that there is a slight performance drop in emotion recognition potentially due to the information loss caused by kl-divergence loss in VAE training for factorization, which is a trade-off between disentanglement and reconstruction. This factorization VAE is however a key backbone in achieving our attribute-aligned representation.

To study how adversarial branches work in our framework, we compare LR-VAE results to LR-VAE(w/o adv). It shows that without adversarial learning in explicitly purifying the emotion-related (identity-related) dimension to identity-free (emotion-free), the representation learned is not ``clean'' enough. Hence, while LR-VAE(w/o adv) also achieves competitive results on main tasks, the sensitive attribute-preserving results are usually worse. This also demonstrates that the emotion-related (identity-related) attributes may contains identity (emotion) information if not explicitly cleaned.


\vspace{-2mm}
\subsubsection{Analysis of Aligned Attributes}
\label{sssec:concentration}

In this section, we further discuss the effectiveness of layered dropout that align attribute-specific information to both ends of the latent codes. We conduct an experiment with the following procedure: we encode the input features into latent codes; next, we divide the latent dimensions into 32 groups; then, for each step, we mask one additional group of latent codes, and train two models, one for emotion recognition, and the other for speaker verification. We compare the performance curve of LR-VAE and A-VAE to observe how layered dropout influence the discriminatory power of the chosen latent code dimension.

We first study the privacy-preserving speech emotion recognition task. The results are shown in the upper row of figure \ref{fig:curve}. In this experiment, we start masking from the bottom of the latent code, which contains more identity-related attributes, to the top of the latent code (more emotion-related attributes). As the procedure moves on, the speaker verification performance steadily decreases (EER increases) until the masking process reaches the middle part of the latent code, where it results in a high EER indicating the point where we achieve an identity-free representation. We can also see that EER curves of LR-VAE and A-VAE intersects, which shows that the masked LR-VAE latent can better eliminate the identity-related attributes. On the other hand, we also observe that the emotion recognition performance slightly decreases toward the ending portion of masking process due to a significant reduction in the node dimension, though A-VAE has an even earlier performance drop.

Next, we study the emotion-protection speaker verification task. The results are shown in the lower row of figure \ref{fig:curve}. In this part of experiment, we start masking from the top of the latent code, similar to the previous procedure, but in a reverse order. As the progress moves on, the emotion recognition performance steadily decreases (weighted f-score decreases), and finally reaches to a similar result comparing to A-VAE. On the other hand, we can see that the speaker verification performance of LR-VAE is better preserved comparing to A-VAE, i.e., the EER curve of LR-VAE is lower in the beginning and increases slower comparing to A-VAE.

\vspace{-2mm}
\section{Conclusions and Future works}

In this paper, we propose a novel disentangled layered speech representation learning that can flexibly preserve sensitive attribute in a unified single training architecture. Compared with other methods, our method achieves a competitive performance on identity-free SER and an improvement on emotionless SV. Also, we show that our proposed method help in pushing the emotion and identity information toward the both ends of the latent codes, and this strategy provides a flexible mechanism to select the target sensitive attributes to protect. Moreover, our attribute aligned learning strategy reduce the training and memory cost as we require only single process and single model to achieve competitive privacy-preserving results on SER and SV against adversarial training, which requires training twice and two models.

In the future, we will generalize our attribute aligned representation from two specific task to general multi-attributes scenarios. We could utilize the middle portion of the latent codes to capture other information about the speaker, e.g., gender, personality, semantics, etc, in order to provide a more complete profile on this factorized speech representation. Moreover, as the disentanglement achieved by kl divergence loss causes information loss, different factorization methods may be applied to enhance our representation capacity.


\bibliographystyle{IEEEtran}

\newlength{\bibitemsep}\setlength{\bibitemsep}{.5\baselineskip plus .05\baselineskip minus .05\baselineskip}
\newlength{\bibparskip}\setlength{\bibparskip}{0pt}
\let\oldthebibliography\thebibliography
\renewcommand\thebibliography[1]{%
  \oldthebibliography{#1}%
  \setlength{\parskip}{\bibitemsep}%
  \setlength{\itemsep}{\bibparskip}%
}
\bibliography{mybib}


\end{document}